\documentclass[3p,times,twocolumn]{elsarticle}

\usepackage{ecrc}
\usepackage{siunitx} 

\volume{00}

\firstpage{1}

\journalname{Nuclear Physics B Proceedings Supplement}

\runauth{R.\,G.\,H.\,Robertson}

\jid{nuphbp}

\jnltitlelogo{Nuclear Physics B Proceedings Supplement}

\usepackage{amssymb}

\usepackage[figuresright]{rotating}

\begin{document}

\begin{frontmatter}




\title{Direct probes of neutrino mass}


\author{R.\,G.\,Hamish Robertson}

\address{Center for Experimental Nuclear Physics and Astrophysics \\ and \\
Dept. of Physics, University of Washington, Seattle WA 98195, USA}

\begin{abstract}
The discovery of neutrino oscillations has shown that neutrinos, in contradiction to a prediction of the minimal standard model, have mass.  Oscillations do not yield a value for the mass, but do set a lower limit of 0.02 eV on the average of the 3 known eigenmasses.  Moreover, they make it possible to determine or limit all 3 masses from measurements of electron-flavor neutrinos in beta decay.  The present upper limit from such measurements is 2 eV.  We review the status of laboratory work toward closing the remaining window between 2 and 0.02 eV, and measuring the mass.

\end{abstract}


\end{frontmatter}


\section{Introduction}
\label{sec:intro}

The mass of the neutrino has been a topic of speculation and research since the  theory of beta decay was  formulated by Fermi \cite{fermi_tentativo_1933}.  Neutrino mass affects the shape of the beta spectrum, and even with the limited data available at the time, Fermi was able to conclude that the mass of the neutrino must be ``either zero, or in any case very small, in comparison to the mass of the electron'' \cite{Wilson:1968zz}.   The discovery by Alvarez and Cornog \cite{PhysRev.56.613} in 1939 that tritium was radioactive and had a small Q-value was important because the effect of neutrino mass is relatively greater in that case.  Moreover, tritium has a simple atomic structure,  a uniquely valuable property as the sensitivity of neutrino mass experiments has advanced over the years.   Seventy-five years later, tritium remains the isotope of choice in the continuing quest to measure the  mass of the  neutrino.  

The first neutrino mass determinations from the shape of the tritium spectrum were carried out with proportional counters in 1948 by Hanna and Pontecorvo \cite{PhysRev.75.983.3} at Chalk River and by Curran {\em et al.}~\cite{PhysRev.76.853} in Glasgow.  Hanna and Pontecorvo were able to set a limit of 500 eV on the neutrino mass, Curran {\em et al.}~about 1 keV.   Experimental work continued until the discovery of parity non-conservation in the weak interaction in 1956 \cite{Lee:1956qn} and the measurement of the helicity of the neutrino in 1958 \cite{Goldhaber:1958nb} made plausible the idea that the neutrino was massless, and  massless neutrinos were subsequently built into the standard model.  A decade-long hiatus  in searches for neutrino mass followed.  A second kind of neutrino, the muon neutrino, was discovered in 1962 \cite{1962PhRvL...9...36D}, and a third, the tau neutrino, was found to exist in 1975 \cite{1975PhRvL..35.1489P}.   Interest in direct searches for neutrino mass revived once it was realized that neutrinos were not responsible for parity violation because the neutral current \cite{Hasert:1973ff} also violated parity \cite{Prescott:1978tm}.  A flurry of excitement pervaded the neutrino-mass community when in 1981 Lyubimov {\em et al.}~\cite{Lyubimov:1980un} reported that the electron antineutrino had a mass of 30 eV, a value that closes the universe gravitationally and would explain its observed flatness.  Unfortunately this result turned out to be incorrect, as was shown by a group at Los Alamos \cite{Robertson:1991vn}.  The Los Alamos experiment used a source of gaseous T$_2$, because Bergkvist had observed in 1972 \cite{Bergkvist:1972xb}  that experiments had reached a sensitivity where it becomes essential to consider molecular and atomic excitations in interpreting the spectrum.   It was almost certainly the complexities of the tritiated valine molecule that contributed to the erroneous result of Lyubimov {\em et al.}

\begin{figure}[h]
\centerline {
\includegraphics[width=3.5in]{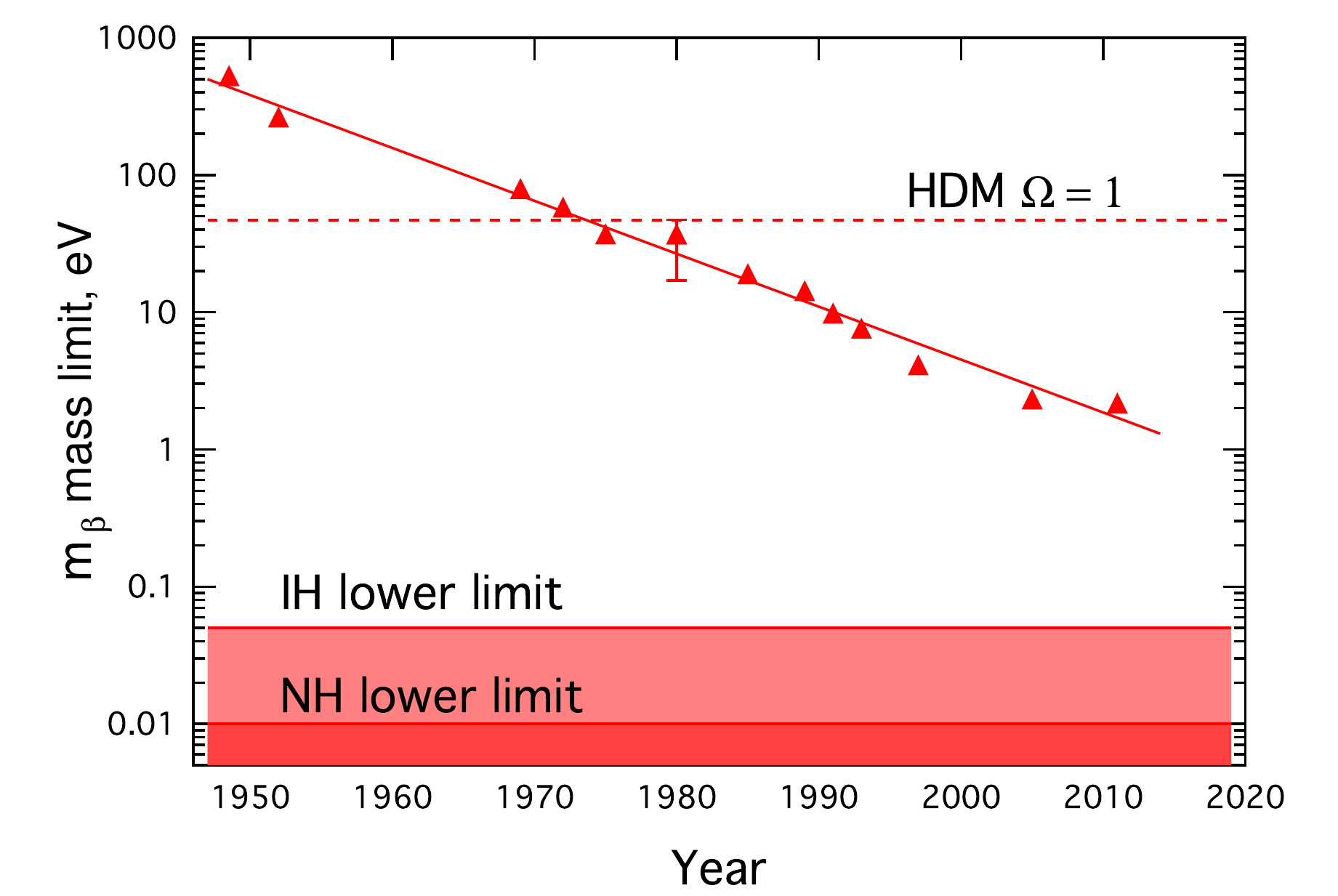}
}
\caption[]{Upper limits on neutrino mass obtained from tritium beta decay experiments {\em vs.} year.  The point with error bars is the non-zero value reported by Lyubimov {\em et al.}~\cite{Lyubimov:1980un}.  Also indicated by HDM  is the mass that would close the universe in the absence of other contributions, and the lower bounds set by oscillations for the inverted and normal hierarchies.
}
\label{fig:mbeta_moores}
\end{figure}

The discovery of neutrino oscillations in atmospheric, solar, and reactor neutrinos \cite{Fukuda:1998mi,Ahmad:2002jz,Eguchi:2002dm} brought a profound change.  Neutrinos were shown  to have mass, and a lower limit on the average mass of the three eigenstates (0.02 eV) has been established.  All three masses are linked by small differences, which means that direct searches can focus on the experimentally most accessible neutrino (the electron antineutrino) and will be able to determine the 3 masses from a single measurement (with a two-fold ambiguity from the presently unknown hierarchy).  The quantity measured in a beta decay experiment is a weighted sum of eigenmasses:  
\begin{eqnarray}
m_\beta^2 &=& \sum_{i=1,3}\left|U_{ei}\right|^2m_{\nu i}^2, \label{eq:mbetadef}
\end{eqnarray}
where the $m_{\nu i}$ are neutrino eigenmasses and ${\bf U}$ is the Maki-Nakagawa-Sakata-Pontecorvo mixing matrix \cite{Beringer:1900zz}.    The absolute lower bound, which neutrino oscillation measurements provide, is $m_\beta \ge 0.01$ eV/c$^2$ for the normal hierarchy, and $m_\beta \ge 0.05$ eV/c$^2$ for the inverted hierarchy.  When the mass is larger than roughly $ 0.1$ eV/c$^2$, the `quasi-degenerate' regime, $m_\beta \simeq m_{\nu1} \simeq m_{\nu2} \simeq m_{\nu3}$.  Figure \ref{fig:mbeta_moores} shows a steady `Moore's Law' decrease in the upper limit on neutrino mass from experiments spanning more than 60 years.

The current upper limit on neutrino mass comes from two tritium experiments, one at Mainz \cite{Kraus:2004zw}, and the other at Troitsk \cite{Aseev:2011dq}.  The experiments lead to a concordant limit of about 2 eV \cite{Beringer:1900zz}.

Neutrino mass is an important question in physics for two reasons.  It is the first definitive contradiction to the minimal standard model.  The standard model is known to be incomplete, lacking things like gravity and dark matter, but until neutrinos were found to have mass, it had never  produced a disagreement with data.   Understanding the origin and patterns of neutrino mass is thus of great interest, because new physics at a very high mass scale may be responsible.  Neutrinos also play a significant role in the formation of the universe.  They are only a small fraction of the dark matter, but because they cool from relativistic to non-relativistic at a recent epoch in the last few billion years, they have influenced large-scale structure.  Quantifying that influence is desirable, and a laboratory measurement would free cosmologists from the need to include neutrino mass in fits to extract other parameters that can only be obtained from astronomical observation.  

In this review the status of experiments now in progress is considered.   There are two experiments on tritium.  The KATRIN experiment in Karlsruhe is in the final stages of construction.  A new scheme, Project 8, is under exploratory development  in Seattle.    There are also ideas about using $^{187}$Re or $^{163}$Ho embedded in microcalorimeters but the Re decay is hindered, calling for large amounts of this costly element, and the Ho spectrum has a complex shape \cite{Robertson:2014zz}.

\section{The KATRIN experiment}

\begin{figure*}[ht]
\centerline {
\includegraphics[width=6in]{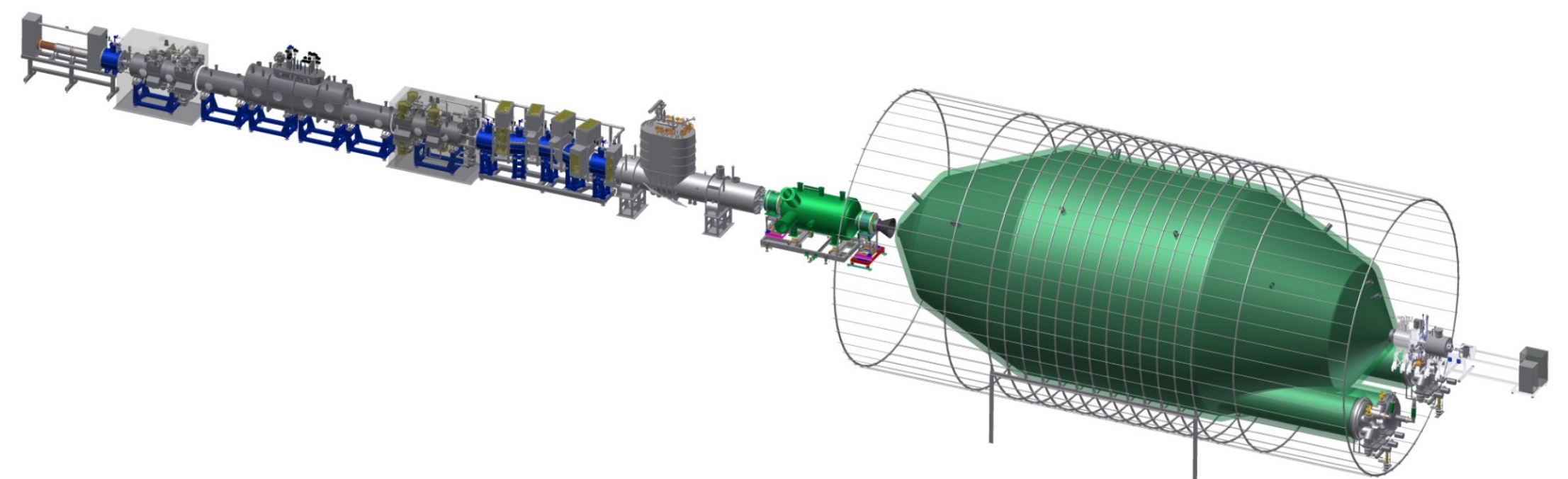}
}
\caption[]{Overview of the KATRIN apparatus. Functional units (from left): Rear system with calibration devices (RS); differential pumping section (DPS1R); windowless gaseous tritium source (WGTS); differential pumping section (DPS1F); differential pumping section with magnetic chicane (DPS2F); cryogenic pumping section with argon frost at 3K (CPS); prespectrometer; main spectrometer; detector.
}
\label{fig:KATRINoverview}
\end{figure*}

The KArlsruhe TRItium Neutrino experiment (KATRIN) couples a gaseous T$_2$ source to a large spectrometer, a `MAC-E' filter based on Magnetic Adiabatic Collimation with Electrostatic retardation.  The experiment is located on the north campus of the Karlsruhe Institute of Technology, contiguous with the prototype tritium-handling facility developed for the ITER controlled fusion experiment now under construction in France.  The apparatus concept is shown in Fig.~\ref{fig:KATRINoverview}.  
The basic operation of the experiment begins with gaseous molecular tritium recirculated through the WGTS at a temperature of 27 K maintained by a two-phase neon cooling loop.   A solenoidal field guides electrons from the source through several stages of pumping and into a prespectrometer that rejects all but the last 100 eV or so of the spectrum.  Electrons that surmount the prespectrometer potential enter the main spectrometer, which has an energy resolution of 0.93 eV base width.  If they also surmount the main spectrometer potential, the electrons are transmitted to a multipixel Si detector for counting.   A spectrum can be built point by point, by stepping either the main spectrometer potential  or the potential of the WGTS.   A comprehensive description of the concept of KATRIN can be found in Ref.~\cite{Angrik:2005ep}.

On site and complete are the prespectrometer, main spectrometer, and detector.   With an electron gun source, the main spectrometer and detector have been undergoing commissioning tests.  The cryogenics for the WGTS have been built and tested: the system performs very well, with temperatures controlled to 4 mK at 27 K \cite{grohmann:2013aa}.  The RS is nearing completion in Santa Barbara.   The DPS1R, DPS1F, and WGTS magnets are built and ready for integration with the cryogenic system.  The monolithic DPS2F system initially constructed suffered a failure of quench protection diodes that were inaccessible and could not be repaired.  A new system is being built in Karlsruhe with 5 separate magnets.  The CPS is under construction in Genoa.   

The main spectrometer and detector operate largely as expected and required for a successful experiment.  The spectrometer has run at up to 35 kV, has delivered sub-eV energy resolution, and has shown background rates of a fraction of 1 c/s. One surprise was the discovery of $^{219}$Rn decaying in the active volume of the spectrometers \cite{Frankle:2011xy}.  This unusual isotope  in the $^{235}$U series is emanating  from the Zr-rich getter strips used to pump hydrogen isotopes, water, and air.   Installation of liquid-nitrogen-cooled baffles has greatly reduced this background but necessitated a challenging design for high-voltage insulators carrying the LN supply from ground potential.  Another difficulty has been the appearance, during  bakeout, of short circuits in the connections between grid layers that line the shell of the spectrometer.  The grids suppress electron backgrounds at the wall, and function best when operated at potentials differing by a few hundred volts.   Progress has been made in repairing the shorts, and work continues apace.

It is expected that initial operation with tritium could begin in late 2016.  The sensitivity of KATRIN is an upper limit of 0.2 eV at 90\% CL after 5 calendar years, or a discovery at $5\sigma$ of 0.35 eV.

\section{Project 8}

As Bergkvist pointed out in 1972, the final-state spectrum in the decay of  tritium bound in a molecule imposes a line-broadening that must be folded with the theoretical beta spectrum.  The line-broadening exacts both  statistical and  systematic penalties when small neutrino mass distortions are being searched for.  The statistical demands are already severe: only $2\times10^{-13}$ of decays populate the last $\Delta E =1$ eV of the spectrum, and this fraction scales as the cube of $\Delta E$.   The KATRIN experiment reaches its ultimate sensitivity with approximately equal contributions of statistical and systematic uncertainty.  

A new approach (dubbed `Project 8') was proposed by Monreal and Formaggio in 2009 \cite{Monreal:2009za}.  The statistical limit mentioned might be circumvented by not extracting and analyzing electrons from the tritium source gas, but instead detecting the cyclotron radiation they emit while moving in a uniform magnetic field $B$.   The electrons from tritium beta decay are mildly relativistic ($\gamma=1.035$) and the cyclotron frequency depends on the kinetic energy $K$ of the electron:
\begin{equation}
\label{eq:cyclotron}
f_\gamma  \equiv  \frac{f_c}{\gamma} =
\frac{e B}{\gamma m_e},
\end{equation}
\noindent where $e \left(m_e\right)$ is the electron charge (mass), $c$ is the
speed of light in vacuum, and $\gamma=\left(1+K/m_e c^2\right)$ is the Lorentz
factor. The nonrelativistic
frequency $f_c$ is \SI{2.799249110(6)E10}{\hertz\per\tesla}. The orbiting electron emits coherent
electromagnetic radiation with a power spectrum centered at
$f_\gamma$.
Due to the $K$ dependence
of $f_\gamma$, a frequency measurement of the cyclotron radiation is
sensitive to the energy of the electron, and thus provides a new form of  spectroscopy, Cyclotron Radiation Emission  Spectroscopy (CRES).  The statistical advantage is two-fold: because the source gas is transparent to the radiation, a large, high-activity source can be used, and because a range of frequencies can be collected and analyzed at once, the time to accumulate a spectrum is shorter than with a  spectrometer that records data point-by-point.

An experimental verification of this concept being desirable, apparatus was set up at the University of Washington to search for cyclotron emission from single electrons orbiting in a uniform magnetic field.  Somewhat surprisingly, no observation of this has  been reported.  Cyclotron motion and radiation is, of course, well established, but the radiation detected was either from large numbers of electrons or, where single electrons were being studied, involved detection through coupling to their axial motion in a Penning trap \cite{PhysRevLett.100.120801}.

The power radiated by an electron is given by the Larmor formula,
\begin{equation}
\label{eq:power}
P\left(\gamma,\theta\right)=\frac{1}{4\pi\epsilon_0}
\frac{2}{3}\frac{e^4}{m_\mathrm{e}^2 c} B^2 \left(\gamma^2 - 1\right)\sin^2\theta,
\end{equation}
\noindent where $\epsilon_0$ is the permittivity of free space and $\theta$
is the pitch angle of the electron, defined as the angle between the momentum
vector of the electron and the magnetic field. An electron with energy
near the \SI{18.6}{\keV} endpoint of tritium and a pitch angle near $90^{\rm o}$ radiates
approximately 1.2 fW  in a \SI{1}{\tesla} magnetic field.  More power is available at higher fields, but the electron is losing energy more quickly by radiation and the receiver technology is more difficult.  At lower fields, on the other hand, the signal-to-noise ratio with respect to background thermal radiation becomes more difficult. 

The experiment \cite{Asner:2014cwa} was carried out not with tritium but with $^{83m}$Kr, a 1.8-hr isomeric activity that is the daughter of a much longer-lived parent, 86-d $^{83}$Rb.  Not only is the Kr non-contaminating, but it produces a spectrum of nearly monoenergetic conversion electrons at kinetic
energies of \SIlist{17830.0(5); 30227(1); 30424(1); 30477(1)}{\eV}~\cite{Picard92}.

The Kr was introduced into a cell made of a section of WR42 waveguide enclosed with 25-$\mu$m Kapton windows.  A vacuum of order 10 $\mu$Pa was maintained by getter pumps.  The cell was connected by a 1-m length of waveguide to a pair of cascaded low-noise HEMT amplifiers maintained at a physical temperature of 50 K by a Gifford-McMahon cryocooler.  The cell temperature was held at 130 K to prevent Kr from freezing out.  The system noise temperature was measured to be 145 K. A uniform magnetic field of 0.9467(1) T was provided by a warm-bore superconducting magnet, and a weak harmonic trapping field was superimposed on the uniform field by means of a circular copper coil wound on the cell.  This trapping field was necessary in order to keep electrons in the cell long enough to make a precise measurement of their frequencies.   Signals were down-converted through two stages of double-balanced mixers and amplifiers to a 125-MHz wide slice that was digitized by a free-running 250-MSPS digitizer.  

Spectrograms were formed from the digitized data by successive Fourier transforms every 32.8 $\mu$s. The frequency axis was binned in 30.52-kHz bins.  One expects a signal from an electron to show a sudden onset of RF power at a certain frequency at the moment of the decay followed by steady emission that drifts monotonically up in frequency as the electron radiates.  Eventually the electron will collide with a background gas atom and be ejected from the trap.   

\begin{figure}[ht]
\centerline {
\includegraphics[width=3.5in]{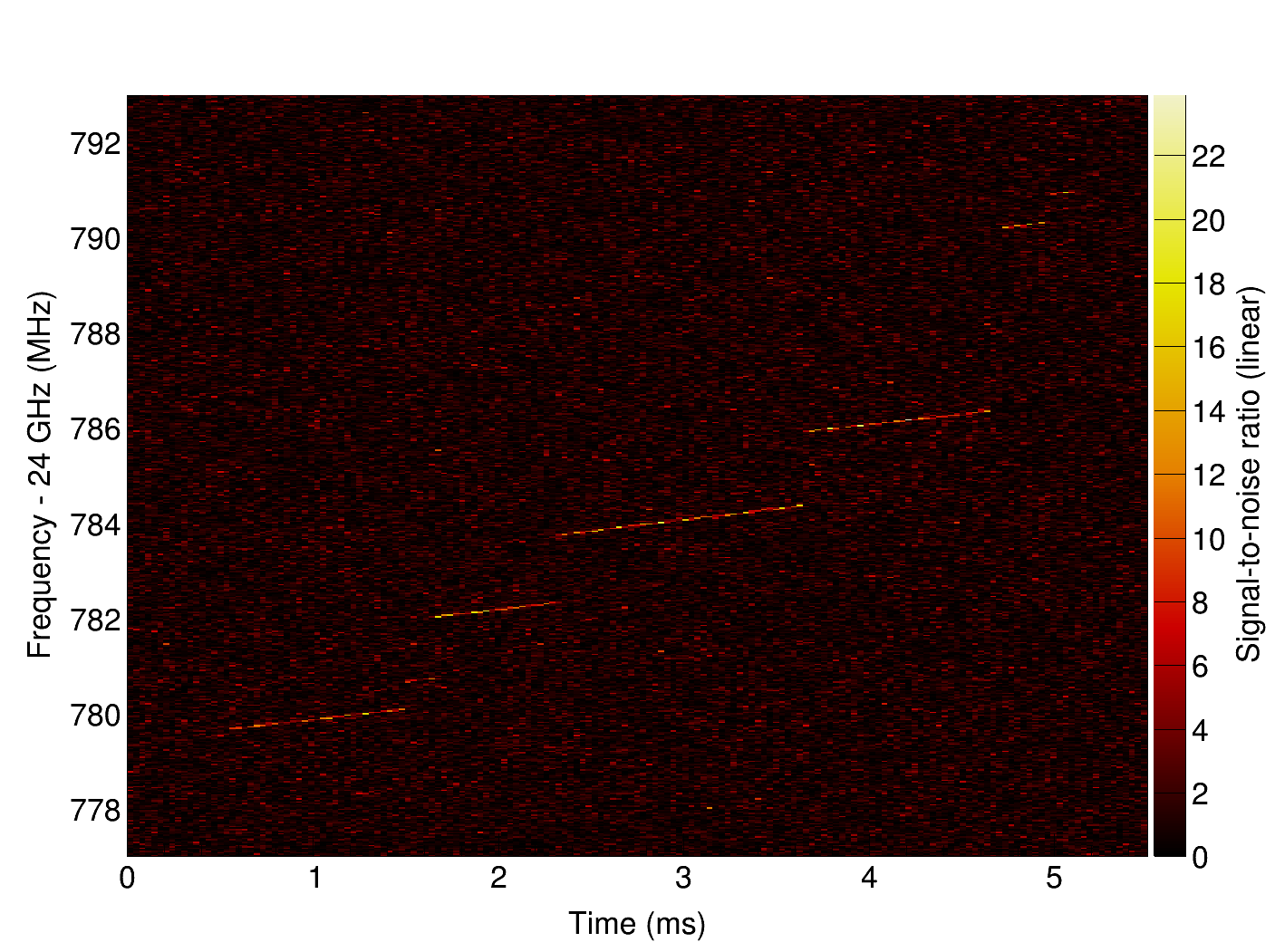}
}
\caption[]{Spectrogram of cyclotron radiation from a trapped 30-keV electron. 
}
\label{fig:candidate0}
\end{figure}

After a number of attempts success was achieved in June 2014.  A spectrogram from the first second of data-taking is shown in Fig.~\ref{fig:candidate0}.
The tracks visible in this spectrogram have the predicted characteristics: a sudden onset of power followed by a slow, quasi-linear increase in frequency.  What was not expected was the persistence of the electron in the trap.  Several collisions with background gas atoms are seen, leading to jumps in the frequency, but the electron is not ejected until the seventh or perhaps eighth collision.  Evidently the changes in pitch angle  in the collision of a 30-keV electron with a gas molecule are typically very small.   A histogram of the frequency jumps from many such events shows that the most probable energy loss is 14 eV, consistent with the expected background gas being hydrogen.
 
\begin{figure}[ht]
\centerline {
\includegraphics[width=3.25in]{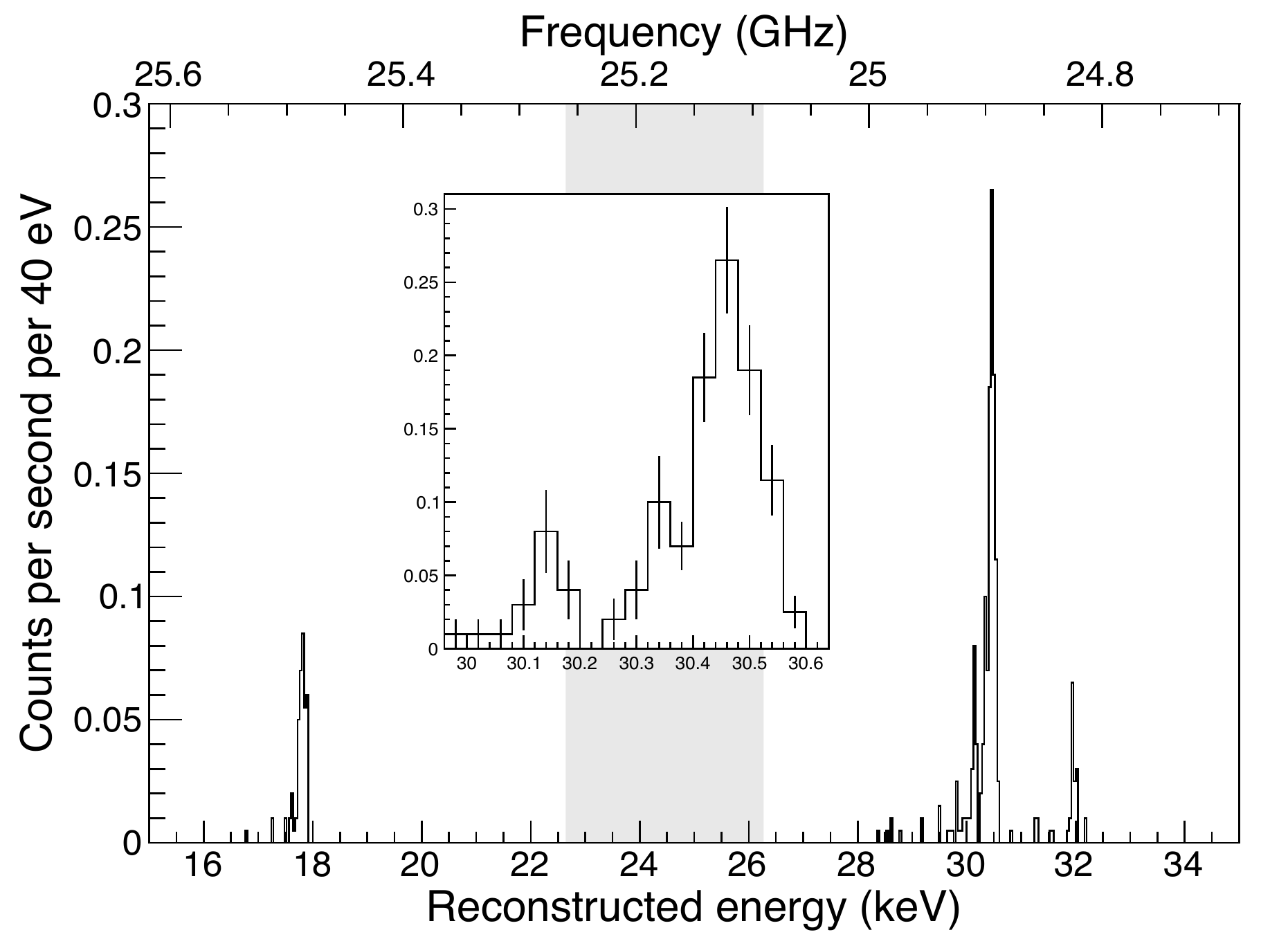}
}
\caption[]{Spectrum of conversion lines from $^{83m}$Kr.  The inset is an expanded view of the 30-keV line.  The energy resolution in this spectrum is about 160 eV.   (The 30-keV line is in fact a 53-eV doublet.)  The shaded region was not included in the frequency range explored.
}
\label{fig:energy_spectrum}
\end{figure}

An energy spectrum was formed by analyzing the tracks to find the lowest frequency in each event, the frequency at the instant the electron was emitted.  The spectrum is shown in Fig.~\ref{fig:energy_spectrum}.

It is clear from these results that the basic scheme outlined by Monreal and Formaggio is a viable one for measuring the beta spectrum of tritium.   A plan for further development of this technique includes a small-scale tritium experiment with apparatus not greatly different from the one used for the conceptual verification, and then larger-scale experiments with increasingly sensitive capabilities.  While the statistical advantages of a large source and spectral efficiency may be within reach, the fundamental limitation to experiments with molecular T$_2$ is the final-state distribution, which has a standard deviation of about 0.4 eV.  The distribution is calculable to high accuracy (of order 1\%, although difficult to quantify) but it nevertheless comes at a high price in statistical precision.     For these reasons, exploratory work on an atomic tritium source has begun.  Some encouragement that this might be possible comes from the success of the ALPHA experiment trapping antihydrogen at CERN \cite{Andresen:2011zz}.   The technique involves axial trapping by solenoidal coils, and radial trapping with Ioffe bars on the surface of a cylinder around the source volume.  Electrons require only the axial trap but atoms must also be radially confined.   For atoms with a magnetic moment of 1 Bohr magneton to be trapped, fields of order \SI{5}{\tesla}  are needed and the atoms must be cooled to sub-K temperatures.  If this can be accomplished, the molecular component present as a contaminant will be frozen out.  Very pure atomic tritium is required because the endpoint of the molecular spectrum is 8 eV higher than the atomic spectrum.

\section{Conclusions}

A direct measurement of the mass of the neutrino from a study of the phase space near the endpoint of a beta spectrum is very attractive for a number of reasons.  There is no dependence on whether the neutrino is a Dirac or Majorana particle.   There is no need to know a nuclear matrix element precisely, it serves only as a normalization constant for the count rate.  No complex phases cloud the interpretation of Eq.\ref{eq:mbetadef}.  No cosmological degrees of freedom are correlated with it.  

It is nevertheless an extremely difficult endeavor to study and quantify a region of the beta spectrum that is populated only with a fractional intensity of  $10^{-13} - 10^{-19}$.  Two experimental campaigns are in progress now to reach below the presently known 2 eV level of mass sensitivity toward the lower limit set by oscillations, 0.01 eV.   If 0.05 eV is reached without seeing a signal, then the hierarchy must be normal.  The KATRIN project, by far the largest and most complex tritium beta decay experiment ever conceived, is moving steadily to first tritium operation in 2016 and will eventually either see the mass or limit it to 0.2 eV.  A new idea, Project 8, has just passed its proof-of-concept milestone, and design work on the next phases is beginning.

\section{Acknowledgments}

Preparation of this report has been supported by the U.S. Department of Energy Office of Science, Office of Nuclear Physics under Award Number DE-FG02-97ER41020.

%




\bibliographystyle{elsarticle-num}
\bibliography{testbiba}







\end{document}